\documentclass[conference]{IEEEtran}
\IEEEoverridecommandlockouts

\usepackage{cite}
\usepackage{amsmath,amssymb}
\usepackage{graphicx}
\usepackage{booktabs}
\usepackage{url}
\usepackage{xcolor}
\usepackage[hidelinks]{hyperref}
\usepackage{subcaption}
\usepackage{tikz}
\usetikzlibrary{positioning,arrows.meta,fit,calc}

\begin{document}
\bstctlcite{IEEEexample:BSTcontrol}

\title{A First Look at Starlink Maritime on Cruise Ships}

\author{
  \IEEEauthorblockN{
    Jinwei Zhao\textsuperscript{*},
    Kousar Malekinasab\textsuperscript{*},
    Jianping Pan
  }
  \IEEEauthorblockA{
    University of Victoria \\
    Victoria, BC, Canada \\
    \{clarkzjw, kovvy, pan\}@uvic.ca
  }
}

\maketitle

\begingroup
\renewcommand{\thefootnote}{\fnsymbol{footnote}}
\footnotetext[1]{Both authors contributed equally to this work.}
\endgroup

\begin{abstract}
Low-Earth-Orbit (LEO) satellite networks are reshaping maritime connectivity, offering much lower latency and higher capacity than traditional geostationary (GEO) links.
As a single user terminal (UT) cannot serve hundreds to thousands of passengers, a cruise ship often deploys several UTs behind an onboard multi-WAN router for aggregated capacity and load balancing.
Consequently, passenger connectivity depends not only on the satellite links but also on how the router distributes sessions among UTs.
In this paper, we measure Starlink maritime connectivity on two cruise ships of different scales, \textit{Utopia of the Seas} near the Bahamas in December 2025 and \textit{Silver Muse} in the Adriatic Sea in July 2026, combining UT telemetry, \textit{inside-out} measurements from a passenger device, and \textit{outside-in} measurements from cloud servers near Starlink Points of Presence (PoPs).
We find that the onboard network, rather than the satellite links, largely determines the passenger experience, as cruise lines rate limit each passenger device.
Moreover, the public IP address of a passenger device changes not only at PoP handovers but also within the same PoP.
Furthermore, while the country code reported by the UT reflects the maritime jurisdiction of the ship, the serving PoP follows neither this jurisdiction nor geographic proximity alone, suggesting factors beyond the distance-based PoP boundaries observed in Starlink aviation.
To the best of our knowledge, this is the first systematic study of Starlink on commercial cruise ships, complementing prior land mobile and in-flight measurements.

\end{abstract}

\begin{IEEEkeywords}
LEO satellite networks, Starlink, maritime, measurement
\end{IEEEkeywords}

\section{Introduction}

Low-Earth-Orbit (LEO) satellite networks, such as SpaceX Starlink and Eutelsat OneWeb, have reshaped Internet connectivity at sea~\cite{StarlinkBusinessMaritime,SatelliteConnectivitySolutionsa}.
As the leading LEO provider, Starlink is increasingly adopted by cruise lines to replace their geostationary (GEO) links, which suffer from hundreds of milliseconds of latency and a limited capacity shared among hundreds to thousands of passengers on board.
To provide sufficient capacity, a cruise ship can choose among three options that differ in scale and dedicated resources.
First, it can install a dedicated \textit{Starlink Bonded Gateway} (previously known as ``Community Gateway''), which scales up to 20~Gbps of symmetric throughput with additional antennas~\cite{StarlinkTransita}.
Royal Caribbean, for instance, has deployed one with 10~Gbps on \textit{Star of the Seas}~\cite{StarlinkProgressReport_2025}.
However, it requires a group of four or more dedicated parabolic gateway antennas and a Transit Service subscription, with pricing available only on request~\cite{WhatStarlinkDedicated}.
Second, \textit{Bonded Starlinks} groups six or more Starlink Performance user terminals (UTs) under a Dedicated Service subscription, providing up to 2.4~Gbps downlink and 600~Mbps uplink with a dedicated Ku-band beam allocation that reduces contention with other users in the same service cell~\cite{WhatStarlinkDedicated}.
Third, and most commonly, a ship deploys multiple off-the-shelf UTs on standard maritime subscriptions behind an onboard multi-WAN router such as Peplink~\cite{MultiStarlinkBondingPeplink}, where the UTs may share the capacity of each service cell with other users, especially near the coast and in ports.

Starlink maritime connectivity on a cruise ship differs from that of residential, land mobile, and aviation users in both mobility and scale.
First, a ship travels for days across territorial seas and international waters, yet moves slower than a land vehicle and far slower than an aircraft, which crosses the boundaries between Starlink Points of Presence (PoPs) within hours.
At open sea, a large cruise ship also offers a clear sky and a stable platform, unlike a land vehicle that frequently encounters transient obstructions or an aircraft whose terminals must cope with rapid attitude changes and air turbulence~\cite{zhaoMeasuringStarlinkAviation2026,zhaoMobileStarlinkarxiv}.
When berthed, however, a ship may stay for hours next to structures that block part of the sky.
Second, a ship shares its UTs among far more users.
While a residential or vehicle UT serves a few users and a widebody aircraft serves a few hundred passengers with two aero terminals, a cruise ship often serves hundreds to thousands of passengers through an operator-managed network with per-user rate limits.

These differences leave several aspects of Starlink maritime connectivity unexplored.
For a ship that sails close to PoP boundaries or stays berthed for hours, it remains unclear how Starlink associates the ship with its PoPs, whether the public IP address of a passenger device remains stable, and whether the 15-second satellite handover, well studied for fixed users, remains the main source of performance degradation at sea.
For passengers sharing multiple UTs, it is also unclear to what extent the experience of a passenger is determined by the satellite links rather than by the onboard network.
Prior maritime studies measure a single UT near the coast~\cite{hossainPredictiveModelStarlink2026a,liDualtaskPredictionModel2026a,denizerFjordLinkComparisonStarlink2025}, while in-flight studies cannot access UT telemetry, as the gRPC interface is blocked on aircraft~\cite{zhaoMeasuringStarlinkAviation2026,jangGEOLEOFirst2025}.

\begin{figure*}[tb]
  \centering
  \includegraphics[width=0.76\textwidth]{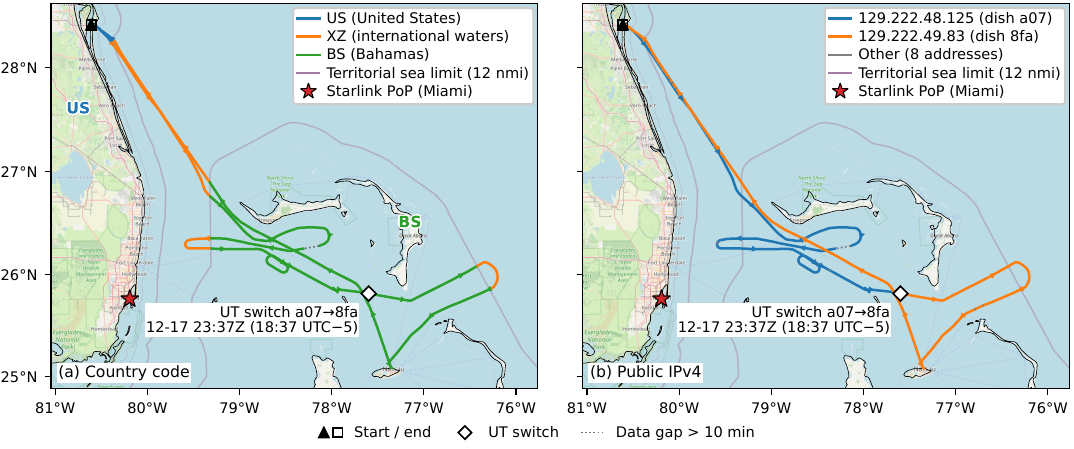}
  \caption{Voyage of \emph{Utopia of the Seas}, December 16--19, 2025: (a) country code and (b) public IPv4 address.}
  \label{fig:utopia-gps-tracks}
\end{figure*}

\begin{figure*}[tb]
  \centering
  \includegraphics[width=0.76\textwidth]{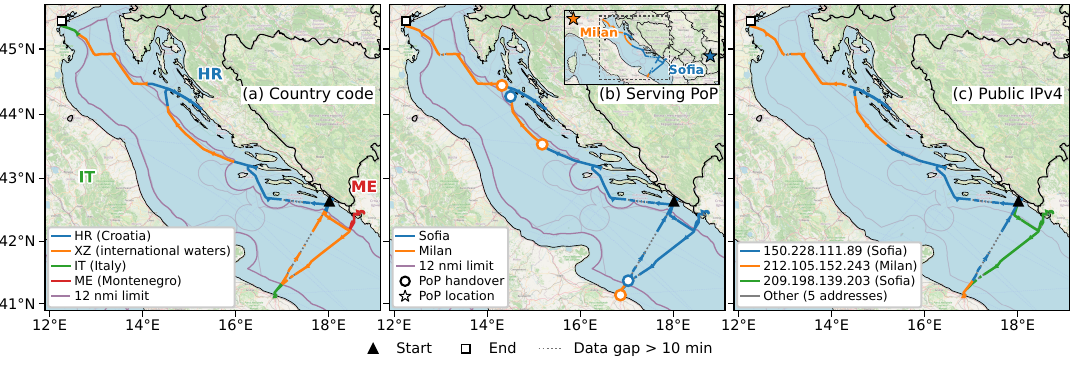}
  \caption{Voyage of \emph{Silver Muse}, July 6--12, 2026: (a) country code, (b) serving PoP, and (c) public IPv4 address.}
  \label{fig:silvermuse-gps-tracks}
\end{figure*}

To address these gaps, we measured two cruise ships of different scales with the same methodology: \textit{Utopia of the Seas} on a voyage from Port Canaveral, Florida, to the Bahamas in December 2025 (Fig.~\ref{fig:utopia-gps-tracks}), and \textit{Silver Muse} for 132 hours during a voyage in the Adriatic Sea in July 2026 (Fig.~\ref{fig:silvermuse-gps-tracks}).
On each ship, a single passenger device that stayed in a cabin recorded the UT status, GPS location, and outage history through the gRPC interface of the Starlink UT, looked up the public IP address, and measured the round-trip time (RTT) to the Starlink gateway at the PoP.
Meanwhile, cloud servers near the serving PoPs measured the RTT to the public IP address of the UT, providing an independent \textit{outside-in} view that excludes the onboard Wi-Fi.
Our main contributions are as follows.

\begin{itemize}
\item We find that the country code reported by the UT reflects the maritime jurisdiction of the ship, whereas the serving PoP follows neither this jurisdiction nor the distance to the PoPs alone, and can change even at berth.
\item We show that, without the ``anycast users'' feature, the public IP address of a passenger device changes not only at PoP handovers but also within the same PoP, when the onboard router moves the device to another UT or even when neither the serving PoP nor the UT changes.
\item We show that the onboard network, rather than the satellite links, largely determines the passenger experience, as it adds about 8--9~ms of latency at the median and a much longer tail, while per-user rate limits cap the throughput of a passenger device. With access to UT telemetry, which is blocked on aircraft~\cite{zhaoMeasuringStarlinkAviation2026}, we also show that the \textit{outside-in} RTT closely matches the UT-reported PoP latency.
\item We find that most outages at sea are brief, obstructed outages concentrate in ports, and an outage can last until the next scheduled satellite handover, yet an outage of one UT does not necessarily interrupt a passenger flow.
\end{itemize}

\section{Related Work}
\label{sec:related}

\emph{Starlink measurement.} Starlink has been studied extensively from stationary UTs under the control of the researchers, covering its access network and satellite links~\cite{panMeasuringLowEarthOrbitSatellite2023d,panMeasuringSatelliteLinks2024d}, transport~\cite{jiangStarTCPHandoverawareTransport2024,kamelStarQUICTuningCongestion2024c}, application performance~\cite{ramanDissectingPerformanceSatellite2023b,mohanMultifacetedLookStarlink2024b}, and large-scale \textit{outside-in} probing~\cite{izhikevichDemocratizingLEOSatellite2024c,wangLargeScaleIPv6BasedMeasurement2026a}.
For mobile users, including in-flight connectivity~\cite{jangGEOLEOFirst2025,zhaoMeasuringStarlinkAviation2026}, the PoP assignment can be approximated by a distance-based Voronoi diagram~\cite{wangCrossUSViewStarlinks2026,zhaoMeasuringStarlinkAviation2026}, although the measured PoP boundaries deviate notably from geographic proximity in places~\cite{zhaoMeasuringStarlinkAviation2026}.
A cruise ship, however, spends many hours in ports, does not keep its public address across PoPs, and shares several UTs among its passengers behind a router they cannot control.

\emph{Maritime connectivity.} Hossain \emph{et al.} measured a fixed UT at coastal sites on land in Texas for throughput prediction~\cite{hossainPredictiveModelStarlink2026a}, which Li \emph{et al.} extended to latency~\cite{liDualtaskPredictionModel2026a}, and Denizer \emph{et al.} compared Starlink and 5G on a research vessel in coastal waters, finding similar Starlink performance at sea and on land~\cite{denizerFjordLinkComparisonStarlink2025}.
These studies focus on the end-to-end performance of a single UT under the control of the researchers.
To the best of our knowledge, no prior work has systematically studied Starlink maritime connectivity on commercial cruise ships, where multiple UTs behind a multi-WAN router serve thousands of passengers and the PoP association and handover of the ship remain unexplored.

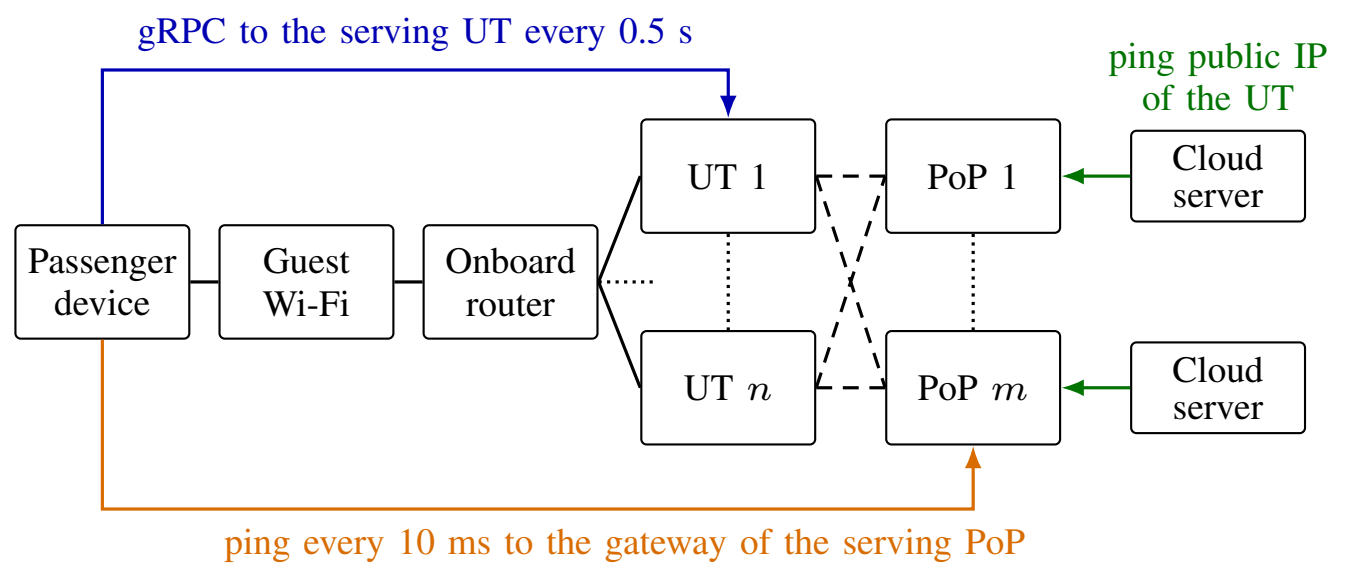
\begin{figure}[t]
\centering
\begin{tikzpicture}[font=\scriptsize, x=0.9mm, y=1mm,
  box/.style={draw, rounded corners=1pt, align=center, inner sep=1.5pt, minimum height=7.5mm, minimum width=11.5mm},
  probe/.style={-{Latex[length=1.5mm]}, semithick},
  link/.style={-, semithick}]
\node[box] (dev)  at (0,0)   {Passenger\\device};
\node[box] (wifi) at (15,0)  {Guest\\Wi-Fi};
\node[box] (rt)   at (30,0)  {Onboard\\router};
\node[box] (ut1)  at (46,7)  {UT 1};
\node[box] (utn)  at (46,-7) {UT $n$};
\node[box] (pop1) at (64,7)  {PoP 1};
\node[box] (popm) at (64,-7) {PoP $m$};
\node[box, minimum height=6mm] (cs1) at (82,7)  {Cloud\\[-1pt]server};
\node[box, minimum height=6mm] (csm) at (82,-7) {Cloud\\[-1pt]server};
\draw[densely dotted, semithick] (ut1.south) -- (utn.north);
\draw[densely dotted, semithick] (pop1.south) -- (popm.north);
\draw[link] (dev) -- (wifi); \draw[link] (wifi) -- (rt);
\draw[link] (rt.east) -- (ut1.west); \draw[link] (rt.east) -- (utn.west);
\draw[link, densely dotted] (rt.east) -- (40.8,0);
\draw[link, densely dashed] (ut1.east) -- (pop1.west); \draw[link, densely dashed] (utn.east) -- (popm.west);
\draw[link, densely dashed] (ut1.east) -- (popm.west); \draw[link, densely dashed] (utn.east) -- (pop1.west);
\draw[probe, green!45!black] (cs1.west) -- (pop1.east);
\draw[probe, green!45!black] (csm.west) -- (popm.east);
\node[green!45!black, align=center, anchor=south] at (cs1.north) {ping public IP\\of the UT};
\draw[probe, blue!70!black] (dev.north) -- ++(0,10.2) -| node[pos=0.25, above]{gRPC to the serving UT every 0.5~s} (ut1.north);
\draw[probe, orange!85!black] (dev.south) -- ++(0,-11.2) -| node[pos=0.3, below]{ping every 10~ms to the gateway of the serving PoP} (popm.south);
\end{tikzpicture}
\caption{Path from a passenger device on a cruise ship to the Starlink PoPs and our measurement setup}
\label{fig:setup}
\end{figure}

\section{Measurement Methodology}
\label{sec:method}

In this section, we describe the measurement methodology applied on both ships, as shown in Fig.~\ref{fig:setup}.
A passenger device in a cabin collects UT telemetry and \textit{inside-out} measurements over the guest Wi-Fi, while cloud servers near the serving PoPs collect \textit{outside-in} measurements toward the public IP address of the serving UT.
Table~\ref{tab:ships} summarizes the two voyages.

\subsection{Voyages and Onboard Network}

\textit{Utopia of the Seas} sailed from Port Canaveral, Florida, to the Bahamas and back, and our measurement covered the voyage from December 16, 01:18 to December 19, 12:51 UTC in 2025 (83.5 hours, Fig.~\ref{fig:utopia-gps-tracks}). \textit{Silver Muse} sailed a round trip from Venice, and our measurement started mid-voyage near Dubrovnik and covered the Bay of Kotor, Bari, Dubrovnik, Hvar, Zadar and the return to Venice, from July 6, 17:11 to July 12, 04:54 UTC in 2026 (131.7 hours, Fig.~\ref{fig:silvermuse-gps-tracks}), with stays of at least 9 hours at each port of call.
\textit{Utopia of the Seas} and \textit{Silver Muse} sailed at a median speed of 15 and 11~knots, reaching up to about 25 and 16~knots, respectively.

A passenger device joins the guest Wi-Fi through a captive portal and stays in the cabin throughout the voyage, so the UT that carries its traffic depends mainly on the load balancing policy of the onboard router.
On both ships, the local network does not support IPv6, and the open HTTP port of the local gateway identifies it as a Peplink router.
Unlike on Starlink-equipped aircraft~\cite{jangGEOLEOFirst2025,zhaoMeasuringStarlinkAviation2026}, the gRPC interface of the UTs was accessible from the local network on both ships.
It shows that the UTs are flat high-performance UTs (\texttt{hp1\_proto0} or \texttt{hp1\_proto1}) with a \texttt{BUSINESS\_PLUS} subscription.
The gRPC interface also exposes the GPS location of the UT.

\subsection{Inside-out Measurements}

\subsubsection{UT telemetry}
The passenger device records the UT status and GPS location through the gRPC interface every 0.5 seconds.
As the router distributes the gRPC session among the UTs, the dish ID and hardware revision identify the UT behind it.
On \textit{Utopia of the Seas}, the gRPC session reached 11 UTs in total, but two of them carried most of our traffic, while the others appeared in only a few responses.
On \textit{Silver Muse}, the gRPC session reached 3 UTs in total, one of which handled 99.94\% of the requests, while the other two appeared only briefly around PoP handovers.
The UT also exports its outage history with the cause, start timestamp, and duration of each outage.

\subsubsection{Public IP address and serving PoP}
The device looks up its public IP address about every 5 seconds and maps it to a serving PoP by the longest prefix match against the Geofeed and PoP assignments published by Starlink~\cite{zhaoMeasuringStarlinkAviation2026}.
Our \texttt{traceroute} measurements further show that each UT has a public IPv4 address without carrier-grade network address translation (CGNAT), which makes it reachable from the Internet.
During our measurements, the UTs on \textit{Utopia of the Seas} were associated only with the Miami PoP (\texttt{mmmiflx1}), and those on \textit{Silver Muse} only with the Sofia (\texttt{sfiabgr1}) and Milan (\texttt{mlnnita1}) PoPs.

\begin{table}[t]
\vspace*{3pt} % keep caption below EDAS 0.7in top margin
\centering
\caption{Measurement Overview}
\label{tab:ships}
\begin{tabular}{lcc}
\toprule
 & \emph{Utopia of the Seas} & \emph{Silver Muse} \\
\midrule
Passenger capacity & 5,668~\cite{UtopiaSeasSize} & 632~\cite{SilverMuseLuxury2027} \\
Sailing region & Bahamas & Adriatic Sea \\
Measurement period & Dec 16--19, 2025 & Jul 6--12, 2026 \\
Measurement hours & 83.5 & 131.7 \\
Serving Starlink PoPs & Miami & Sofia, Milan \\
Passenger device & Mini PC & Laptop (VM) \\
Throughput tests & -- & iPerf3 \\
\bottomrule
\end{tabular}
\end{table}

\begin{figure*}[tb]
  \centering
  \begin{subfigure}[h]{0.32\textwidth}
    \includegraphics[width=\columnwidth]{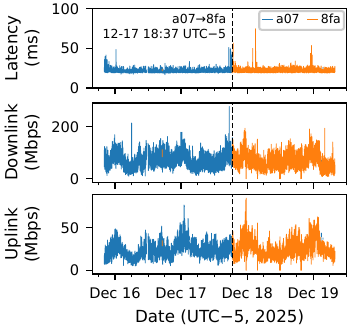}
    \caption{\textit{Utopia of the Seas}}
  \end{subfigure}
  \begin{subfigure}[h]{0.32\textwidth}
    \includegraphics[width=\columnwidth]{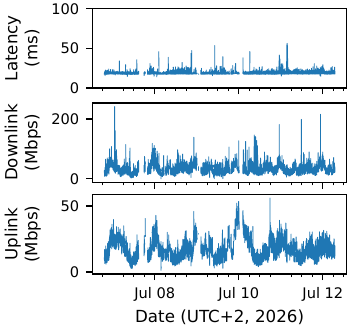}
    \caption{\textit{Silver Muse}}
  \end{subfigure}
  \begin{subfigure}[h]{0.32\textwidth}
    \includegraphics[width=\columnwidth]{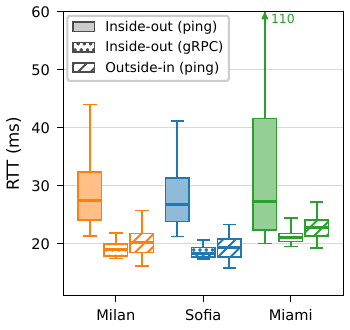}
    \caption{RTT by serving PoP}
  \end{subfigure}
  \caption{UT telemetry on both ships and RTT to the serving PoP.}
  \label{fig:utopia-time-series}
\end{figure*}

\begin{figure}[tb]
  \centering
  \includegraphics[width=0.7\columnwidth]{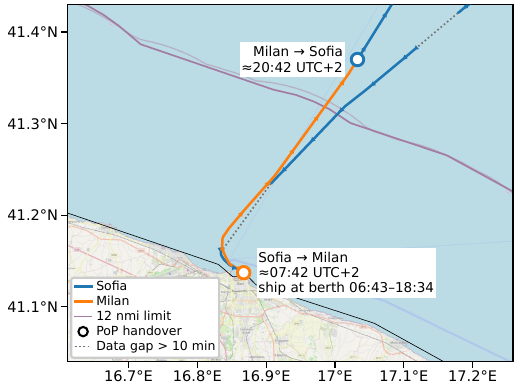}
  \caption{PoP handovers of \emph{Silver Muse} in Bari, switching to Milan while berthed and back to Sofia after departure.}
  \label{fig:silvermuse-pop-handover}
\end{figure}

\subsubsection{Latency}
Instead of the CGNAT gateway \texttt{100}.\texttt{64}.\texttt{0}.\texttt{1} used for residential users~\cite{panMeasuringLowEarthOrbitSatellite2023d}, the device sends ICMP pings every 10~ms to the Starlink gateway of its current public IP prefix, e.g., \texttt{209}.\texttt{198}.\texttt{138}.\texttt{1} for \texttt{209}.\texttt{198}.\texttt{138}.\texttt{0}/\texttt{23}.

\subsubsection{Throughput}
On \emph{Silver Muse}, we additionally measured the uplink and downlink throughput between the passenger device and a cloud server near the serving PoP with iPerf3 over a single TCP stream, using both the BBR and CUBIC congestion control algorithms.
Each test lasts 60 seconds, followed by a 60-second idle period.
We did not run systematic iPerf3 tests on \emph{Utopia of the Seas}.
On both ships, the gRPC interface also reports the throughput of the UT, which aggregates the traffic of all passengers it serves.

\subsection{Outside-in Measurements}

Cloud servers near the Miami, Sofia and Milan PoPs send ICMP requests every 10~ms to the public IP address of the router attached to the UT serving the passenger device.
Unlike the anycast router addresses of the Starlink aviation service~\cite{zhaoMeasuringStarlinkAviation2026}, the public address changes at a PoP handover and may also change within the same PoP.
We therefore follow the address changes observed by the passenger device and only use the server near the serving PoP toward the current address.
As these ICMP probes terminate at the router, they exclude the onboard network and Wi-Fi.

\subsection{Ethics}

All \textit{inside-out} measurements were collected from our own devices as regular passengers who purchased the onboard Wi-Fi plan.
We did not capture the traffic of other passengers or attempt to bypass the per-user rate limits.
We identified the onboard Peplink router only from the HTTP service of the local gateway and did not attempt to access its management interface.
We used the gRPC interface of the UTs only for read-only requests of their status, location and outage history.
The \textit{outside-in} measurements consist only of ICMP probes toward the public IP address of the ship during the voyages.

\section{Measurement Findings}
\label{sec:results}

In this section, we present our measurement results on the maritime jurisdiction reflected by the country code of the UT, the PoP association and public IP address assignments, the load balancing of the onboard router among UTs, and the latency and throughput of a passenger device.
The gRPC telemetry captures the state of the satellite link, the \textit{inside-out} measurements capture what the passenger device experiences over Wi-Fi, and the \textit{outside-in} probes serve as an independent cross-check that excludes the onboard network.

\subsection{Jurisdiction and PoP Association}

Fig.~\ref{fig:utopia-gps-tracks}(a) and Fig.~\ref{fig:silvermuse-gps-tracks}(a) show the country code reported by the UT along the two voyages.
The country code changes exactly when the ship crosses the 12-nautical-mile (nmi) territorial sea limit.
Within a territorial sea, the UT reports the ISO 3166-1 two-letter code of the coastal state, i.e., US (United States) and BS (Bahamas) for \textit{Utopia of the Seas}, and HR (Croatia), IT (Italy) and ME (Montenegro) for \textit{Silver Muse}, while beyond the limit, it reports XZ, the code that UN/LOCODE assigns to installations in international waters~\cite{DirectoryLOCODE}.

However, the serving PoP changes neither with the country code nor with the distance to the PoPs alone.
For mobile users on land and in the air, the PoP boundaries can be approximated by a distance-based Voronoi diagram, with visible deviations hypothesized to stem from the peering capacity, subscriber density and load balancing of each PoP~\cite{wangCrossUSViewStarlinks2026,zhaoMeasuringStarlinkAviation2026}.
\textit{Utopia of the Seas} was served only by the Miami PoP throughout the voyage, in US and Bahamian waters as well as in international waters, consistent with the observation that the Miami PoP serves a large part of the Caribbean~\cite{zhaoMeasuringStarlinkAviation2026}.
\textit{Silver Muse} was served by the Sofia and Milan PoPs in turn, with five PoP handovers events as shown in Fig.~\ref{fig:silvermuse-gps-tracks}(b).

Fig.~\ref{fig:silvermuse-pop-handover} zooms into the two PoP handovers near Bari, Italy.
The ship berthed at 06:43 local time, and the PoP handover from Sofia to Milan followed at about 07:42 while the ship was still at berth.
The gRPC session briefly reached another UT for tens of seconds, and the original UT kept the Milan address after the session returned, showing that a PoP handover happened for the original UT.
As neither the position nor the jurisdiction of the ship changed, this handover was triggered on the network side.
The PoP handover back to Sofia occurred at about 20:42, two hours after the ship left the port at 18:34 and after it had already crossed the 12-nmi limit into international waters.
These observations suggest that the PoP assignment of Starlink maritime service also depends on network-side factors beyond the position of the ship.

\begin{figure}[tb]
\centering
\includegraphics[width=0.7\columnwidth]{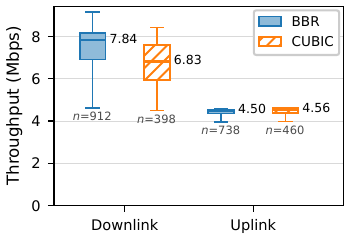}
\caption{TCP throughput of a passenger device on \textit{Silver Muse}.}
\label{fig:silvermuse-throughput}
\end{figure}

\subsection{UT Load Balancing and Public IP Address}

The onboard router distributes passenger sessions among the UTs for load balancing, and the UT serving our device can be inferred from the dish ID in the gRPC response.
On \textit{Utopia of the Seas}, the gRPC session reached 11 UTs in total, but two of them, a07 and 8fa,\footnote{We keep only the last three characters of each dish ID for privacy.} carried most of our traffic, with continuous runs of up to 14.8 and 11.0 hours, respectively.
The serving UT of our device changed from a07 to 8fa on December 17 at 18:37 local time, while the ship was at sea, and 8fa served the device for most of the rest of the voyage as shown in Figs.~\ref{fig:utopia-gps-tracks}(b) and Fig.~\ref{fig:utopia-time-series}(a).

The public IP address can change in three different cases.
First, as neither ship had the ``anycast users'' feature used in Starlink aviation service~\cite{zhaoMeasuringStarlinkAviation2026}, every PoP handover gives the UT a new address from one of the GeoIP prefixes associated with the new PoP.
Second, the address changes within the same PoP when the router moves the device to another UT.
For example, all 10 addresses observed on \textit{Utopia of the Seas} belong to the Miami PoP, and the address of our device changes along with its serving UT.
Third, the address can change even when neither the serving PoP nor the serving UT changes.

\subsection{UT Telemetry and Latency}

Fig.~\ref{fig:utopia-time-series}(a) and Fig.~\ref{fig:utopia-time-series}(b) show the UT telemetry of the two ships in local time.
On both ships, the UT-reported latency to the PoP (\texttt{pop\_ping\_latency\_ms}) has a baseline of about 20~ms throughout the voyage.
The throughput reported by the UT aggregates the traffic of all passengers it serves, and shows a clear diurnal pattern that follows the local time and thus the activity of passengers, with the lowest traffic from midnight to dawn.
The median throughput per UT was 66.6~Mbps downlink and 24.2~Mbps uplink on \textit{Utopia of the Seas}, and 27.6~Mbps downlink and 16.0~Mbps uplink on \textit{Silver Muse}, rising to 132.8~Mbps and 52.0~Mbps, and to 82.8~Mbps and 36.7~Mbps, at the 99th percentile.
Averaged over the busiest 60~s, a single UT sustained up to 162.0~Mbps downlink and 71.2~Mbps uplink on \textit{Utopia of the Seas}, and 139.2~Mbps downlink and 52.4~Mbps uplink on \textit{Silver Muse}.
These values are aggregate statistics of the traffic the UT carried, which reflect passenger demand rather than the capacity of the link.
The higher throughput on \textit{Utopia of the Seas} is likely because it is a much larger ship that serves many more passengers.

Fig.~\ref{fig:utopia-time-series}(c) compares three RTT measurements for each serving PoP.
The \textit{inside-out} ping from the passenger device to the PoP gateway, the UT-reported latency through gRPC, and the \textit{outside-in} ping from the cloud server near the PoP to the public address of the UT.
On \textit{Silver Muse}, the UT-reported PoP latency has a median of 18.3~ms with Sofia and 18.9~ms with Milan, and the \textit{outside-in} RTT has a median of 19.4~ms and 20.1~ms, respectively.
The \textit{outside-in} RTT thus closely matches the PoP latency reported by the UT, with a median only 1.1~ms and 1.2~ms higher.
This small difference comes from the terrestrial paths between the PoPs and the cloud servers in nearby data centers, which only the \textit{outside-in} ping traverses.
In contrast, the \textit{inside-out} ping has a median of 26.7~ms and 27.5~ms and a 95th percentile of 41~ms and 44~ms, respectively.
A passenger device thus experiences about 8--9~ms of additional latency at the median due to the onboard Wi-Fi and local network, along with much higher variation.
On \textit{Utopia of the Seas}, the UT-reported PoP latency and the \textit{outside-in} RTT have medians of 21~ms and 22.8~ms, respectively, while the \textit{inside-out} ping has a median of 27~ms, similar to that on \textit{Silver Muse} but a much longer tail, with a 95th percentile of 110~ms, possibly due to contention on the guest Wi-Fi and congestion in the onboard network.
Therefore, the onboard network, rather than the satellite link, largely determines the passenger experience.
Moreover, consistent with previous studies~\cite{zhaoMeasuringStarlinkAviation2026}, the \textit{inside-out} and \textit{outside-in} latency measurements show the same minimum RTT pattern caused by the scheduled satellite handovers every 15 seconds.
Furthermore, as the \textit{outside-in} RTT closely matches the UT-reported PoP latency on both ships, \textit{outside-in} measurements provide a representative view of the satellite link that excludes the onboard network, a comparison that is not possible on aircraft, where the gRPC interface of the UT is blocked~\cite{zhaoMeasuringStarlinkAviation2026}.

\subsection{Single-User Throughput}

Fig.~\ref{fig:silvermuse-throughput} shows the throughput of our passenger device on \textit{Silver Muse}, measured with iPerf3 using BBR and CUBIC.
The downlink throughput with BBR has a median of 7.84~Mbps and a maximum of 9.38~Mbps, while CUBIC achieves a lower median of 6.83~Mbps, consistent with previous findings that loss-based congestion control such as CUBIC performs worse over Starlink~\cite{jiangStarTCPHandoverawareTransport2024,kamelStarQUICTuningCongestion2024c}.
The uplink throughput is tightly concentrated at a median of 4.50~Mbps with BBR and 4.56~Mbps with CUBIC.
These results suggest that both the downlink and uplink throughput are capped by the onboard router according to the purchased Wi-Fi plan, rather than by the satellite links.
For comparison, the aggregate traffic of a single UT on \textit{Silver Muse} reached 82.8~Mbps downlink and 36.7~Mbps uplink at the 99th percentile.
As hundreds of passengers share a few UTs, such per-user limits help the cruise line manage the shared throughput according to passenger demand and the Wi-Fi plans they purchased.

\begin{figure}[t]
\centering
\centering
\includegraphics[width=0.74\linewidth]{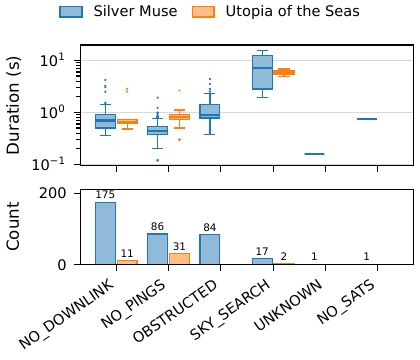}
\caption{Outages reported by the UTs.}
\label{fig:outages}
\end{figure}

\subsection{UT Outages}

Fig.~\ref{fig:outages} summarizes the outage history of the UTs by cause.
Most outages last about 1~s or less, except for \texttt{SKY\_SEARCH}, which lasts several seconds while the UT searches the sky for satellites after a reboot and, in some cases, after a severe \texttt{OBSTRUCTED} event.
We exclude \texttt{BOOTING} outages from Fig.~\ref{fig:outages}, since they come from reboots of the UT rather than outages related to the satellite link.
\textit{Silver Muse} recorded 364 outages in 131.7 hours, far more than the 44 outages of \textit{Utopia of the Seas} in 83.5 hours.
\textit{Utopia of the Seas} recorded no \texttt{OBSTRUCTED} outages, whereas most of the 84 on \textit{Silver Muse} occurred while the ship was berthed at the Port of Dubrovnik in Croatia, next to a bridge that likely blocked part of the line-of-sight of the UT.
From the obstruction maps reported by the UT, we extract the relative SINR and the ID of the connected satellite~\cite{zhaoMobileStarlinkarxiv}, and find that outages of different causes may be part of a single event.
For example, on July 9, 2026, while \textit{Silver Muse} was berthed at the Port of Dubrovnik, the relative SINR dropped sharply at 09:33:49 local time, followed by about 6~s of consecutive \texttt{OBSTRUCTED}, \texttt{NO\_SATS} and \texttt{NO\_DOWNLINK} outages.
Although the UT switched satellites several times through dynamic beam switching~\cite{starlinkStarlinkBeamSwitching,zhaoMobileStarlinkarxiv}, the link did not recover until the next scheduled handover at 09:33:57, when the UT switched to another satellite.

\section{Conclusion}
\label{sec:conclusion}

In this paper, we measured Starlink maritime connectivity on two cruise ships of different scales, combining UT telemetry, \textit{inside-out} measurements from a passenger device, and \textit{outside-in} measurements from cloud servers near the serving PoPs.
We found that the PoP handovers of a ship follow neither its maritime jurisdiction nor the distance-based approximation of PoP boundaries, suggesting additional network-side factors.
The public IP address of a passenger device changes not only at PoP handovers but also within the same PoP, as the onboard router moves the device among UTs.
The onboard Wi-Fi and per-user rate limits are the main bottlenecks of the passenger experience.
Most outages at sea are brief and obstructed outages concentrate in ports, while an outage can last until the next scheduled satellite handover, yet an outage of one UT does not necessarily interrupt a passenger flow.
Further analysis of the impact of obstructions and dynamic beam switching, and of multiple UTs behind an onboard router, such as their aggregated throughput and how load balancing mitigates packet loss and outages, is left to future work.

\noindent\textbf{Acknowledgment:} We thank Lei and Marc for facilitating the measurements on both ships.
This work was supported in part by NSERC, CFI and BCKDF.

\bibliographystyle{IEEEtran}
\bibliography{refs}

\end{document}